# The limitation for popular descriptions of α-relaxation temperature dependence

V. A. Popova and N. V. Surovtsev

*Institute of Automation and Electrometry, Russian Academy of Sciences, Novosibirsk, 630090, Russia*

Applicability of three popular functions (Vogel-Fulcher-Tammann, double activation law and frustration-limited domains model) for the description of the temperature dependence of α-relaxation time $\tau_\alpha$ is considered for three typical glass-formers (propylene carbonate, ethanol and picoline). Two first functions have three free parameters. It was found that while they are in satisfactory agreement with the experimental data of $\tau_\alpha(T)$, they fail in describing the transition from an Arrhenius-like to a non-Arrhenius behaviour. This transition is seen in the derivative analysis of $\tau_\alpha(T)$. We argue that Vogel-Fulcher-Tammann and double activation functions should be applied and compared only at $T < T_A$, where $T_A$ is the temperature of transition from an Arrhenius-like to a non-Arrhenius behaviour. It was shown that the four-parametric frustration-limited domains model with imbedded transition from Arrhenius to non-Arrhenius behaviour at $T = T^*$ also has no advantage in the derivative analysis, since $T^*$ is systematically different from $T_A$ in the cases considered.

Description of the dynamical response of glass-forming materials is still the very popular and interesting topic of condensed matter physics [1-3]. The main relaxation process in glass-forming liquids, α-relaxation, covers the enormously broad range of the relaxation time, $\tau_\alpha$ changing from $10^{-11}$ s in the low-viscous state to $10^3$ s at the glass transition temperature ($T_g$). For the most of glass-forming materials the temperature dependence of their structural α-relaxation dynamics declines from the simple thermoactivated (Arrhenius) law. The challenge to describe the non-Arrhenius temperature behaviour of α-relaxation comes both from a need to empirically describe experimental data and from a wish to provide a trend for glass transition models and theories and ways of their comparison.

At present there are few empirical formulas for the description of the non-Arrhenius temperature behaviour of α-relaxation of glass-forming liquids. The most popular equation is the Vogel-Fulcher-Tammann (VFT) law [4].

$$\tau_\alpha = \tau_0 \exp\left[\frac{B}{T - T_{VF}}\right]. \quad (1)$$

VFT-law provides usually a good description of the experimental data, while in details and/or derivative characteristics some failings can be revealed [5-7]. It is also acceptably, that this equation can be derived in some reasonable models, assuming vacancy type of relaxational motion, which has the linear temperature dependence and disappears at $T = T_{FV}$. The Arrhenius case corresponds to a particular case of equation (1), when $T_{VF} = 0$. Today VFT law is the salient phenomenological description for describing the α-relaxation dynamics of glass-forming liquids.

Else one attractive description was renewed by Mauro *et al* [8]

$$\tau_\alpha = \tau_0 \exp\left[\frac{K}{T}\exp\left(\frac{C}{T}\right)\right]. \quad (2)$$

As VFT-law this equation also contains only three adjustable parameters and corresponds mathematically to double exponential function of $(1/T)$. In analogy with the term of "activation law" for the Arrhenius case, we will refer equation (2) as "double activation law" (DAL), while the effective barrier would origin from entropic effects [8]. It is interestingly to note that Taylor's expansion of $\exp(C/T)$ in equation (2) provides the quadric behaviour of $\log\tau_\alpha$ as the function of $(1/T)$, which is used also in other different description of $\tau_\alpha(T)$ [9,10].

Reference [8] presents the model for configurational entropy based on a constraint approach, which leads to DAL. In certain (maybe vulgar) sense the difference between equation (1) and equation (2) is the change of the temperature dependence for configuration vacancies from a linear to an activation law. This change allows one to avoid the divergence at $T = T_{FV}$ imbedded in VFT-law.

Both empirical functions (VFT-law and DAL) provide rather satisfactory description of $\tau_\alpha$ temperature dependences. However, it is important to find out which function works better. The comparison of these laws was taken on various glass-forming liquids in recent articles [8,11]. Strategy of these works were to compare the experimental data in a maximally broad temperature range, covering the maximally broad range of the relaxation time (extending 16 decades in [11]) or viscosity [8]. According to reference [8] DAL description is more accurate for fit of the experimental results, and in [11] it is concluded that DAL works better in the classical molecular glass, while VFT-law has preferences for alcohols.

The main outlook of the present work is to point out that it is not the best way to compare VFT-law and DAL in a maximally broad temperature range, covering the temperatures above $T_A$, where $T_A$ is the temperature of transition from an Arrhenius-like to a non-Arrhenius behaviour for $\tau_\alpha(T)$. The existence of such transition was illustrated for several glass-forming liquids by the derivative analysis [7,12], where the value $(d\log\tau/d(1/T))^{-1/2}$ versus $(1/T)$ is considered.

Physically $T_A$ would mean the onset of the cooperativity for α-relaxation, at $T > T_A$ rather single molecular response being important. The change of degree of relaxation cooperativity for transition from low-viscous to high-viscous state is intuitively expected and assumed in a number of model approaches [13,14]. On other hand, in some models the appearance of locally favored structures [15,16] or frustration-limited domains [17] is suggested. Raman scattering lineshape analysis [18] and the temperature dependence of the Landau-Placzek ratio [19] reveal the peculiarity at $T = T_A$, naturally interpreting as the appearance of locally favored structures at this temperature. It is believed that these structures plays the important role in preventing crystallization [15,17,20].

The problem is whether one should use VFT-law or DAL at $T > T_A$? None of equations (1) and (2) provide the true peculiarity at $T = T_A$, but both expressions look similar to the Arrhenius law in the high-temperature limit. So, the question is whether VFT-law or DAL with parameter values found from the fit of experimental data would be in satisfactory agreement with the derivative analysis in the spirit of references [7,12]?

We will complete the analysis of VFT-law and DAL by a model of frustration-limited domains (FLD) [17,21]. This model predicts a four-parametric law

$$\tau_\alpha = \tau_0 \exp\left(\frac{E_\infty + BT^*[(T^*-T)/T^*]^{8/3}\Theta(T^*-T)}{T}\right), \quad (3)$$

where $E_\infty$, $B$, $\tau_\infty$ and $T^*$ are temperature-independent adjustable parameters and $\Theta(T^*-T)$ is a step Heaviside function. FLD model uses one parameter more than VFT-law or DAL, it's advantage is that transition at $T = T^*$ is imbedded.

In present work we consider three glass-forming liquids: propylene carbonate, picoline and ethanol. These materials are widely-spread and quite good studied glass-forming liquids. There are extensive experimental data of $\tau_\alpha$ within a wide temperature range for these materials (picoline [22], ethanol [11], propylene carbonate [23]). Experimental data of $\tau_\alpha(T)$ for propylene carbonate, picoline and ethanol were fitted by three expressions considered. All fits were carried out in program package, minimizing mean square error for $\log\tau_\alpha$. The results for the propylene carbonate's data are given in figure 1. Green, blue and red lines are the results of the fits by FLD, VFT and DAL respectively. Qualitatively similar results were obtained for picoline and ethanol liquids. As it is seen from figure 1 all theoretical models provide a rather good fitting quality of the data within the full temperature range. The parameters of the fits are given in table.

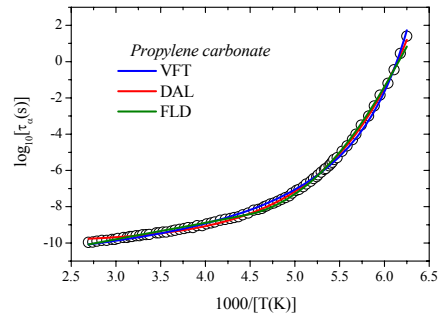

**Fig. 1** Arrhenius plot for $\tau_\alpha(T)$ of propylene carbonate (circles) together with fits by VFT-law (blue line, equation (1)), DAL (red line, equation (2)), FLD model (green line, equation (3)).

| Material | $T_A$, K | VFT, Eq.(1) | | | DAL, Eq (2) | | | FLD, Eq.(3) | | | |
|---|---|---|---|---|---|---|---|---|---|---|---|
| | | $\log\tau_0$ | B, K | $T_{VF}$, K | $\log\tau_0$ | K, K | C, K | $\log\tau_0$ | $E_\infty$, K | B | $T^*$, K |
| PC | 290 | -11.115 | 237 | 142 | -9.944 | 5.24 | 933 | -12.5 | 910.6 | 124 | 231 |
| Ethanol | 172 | -10.516 | 288 | 74 | -9.393 | 54 | 291 | -12.9 | 954.4 | 85.2 | 139 |
| Picoline | 250 | -13.137 | 285 | 111 | -11.82 | 15.6 | 615 | -13,9 | 753.7 | 95.3 | 202 |

**Table** Parameters of the fits of $\tau_\alpha(T)$ and temperature $T_A$ from the derivative analysis.

To visualize the peculiarity of $\tau_\alpha(T)$ dependence in vicinity of $T_A$, the derivative analysis in spirit of [7,12] was used. Let's consider the function $(d\log\tau_\alpha/d(1/T))^{-1/2}$ versus $(1/T)$. This expression linearizes the certain phenomenological descriptions. For this function the Arrhenius law transforms into a constant, while for the VFT-law this function is $(1-T_0/T)B^{-1/2}$ (linear function of $1/T$).

The derivative analysis was performed for propylene carbonate in [23]. The results are presented in the figure 2 by symbols. The transition from a constant law at high temperatures to a linear behaviour at low temperatures is seen in figure 2, indicating temperature $T_A$. It also corroborates the applicability of the VFT law for description of the low-temperature part of relaxation dynamics.

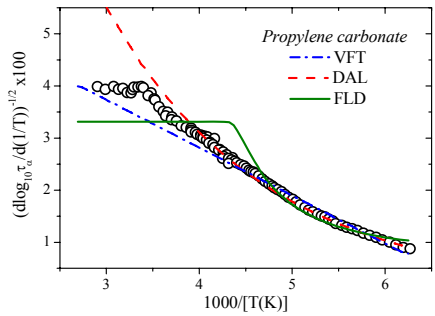

**Fig. 2** The derivative analysis of propylene carbonate (circles, data from [23]) together with results of VFT-law (blue line), DAL (red line) and FLD model (green line).

Similar analysis was performed by us for ethanol [11] and picoline [22] data of $\tau_\alpha$. The result of derivative analysis is given by symbols in figure 3 and figure 4 for ethanol and picoline, respectively (circles, square and stars in the case of picoline correspond to the data from light, neutron and dielectrics spectroscopies, respectively). Phenomenon of a rather sharp transition from the linear-like regime to the constant law as function of $1/T$ is seen for both cases (figures 3 and 4), depicting $T_A$. Such feature in the temperature dependence of the relaxation time seems to be the inherent peculiarity of glass-forming liquids [7,12].

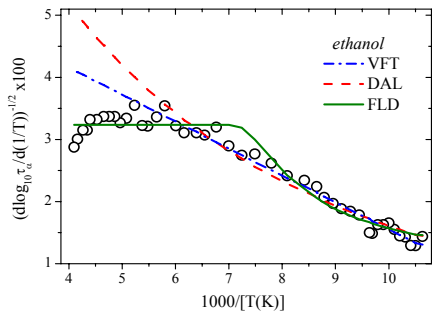

**Fig. 3** The derivative analysis of ethanol (circles, data from [11]) together with results of VFT-law (blue line), DAL (red line) and FLD model (green line).

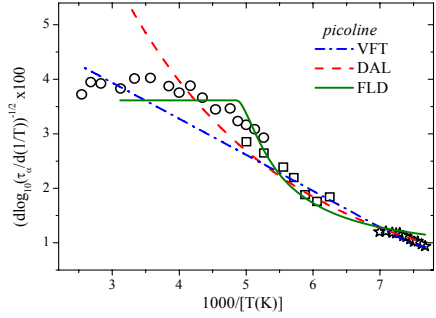

**Fig. 4** The derivative analysis of picoline (symbols, data from [22]) together with results of VFT-law (blue line), DAL (red line) and FLD model (green line).

What about the derivative analysis of the studied fits? As it was noted previously, the VFT law transforms into a lineal behaviour after the derivation. Thus, the derivation of VFT-law works only at temperatures lower $T_A$, being in strong disagreement at $T > T_A$ (figures 2, 3, 4).

It case of DAL model (equation 2) the derivation is

$$(d\log\tau_\alpha/(d1/T))^{-1/2} = \sqrt{K}\exp\left(-\frac{C}{2T}\right)\sqrt{1+\frac{C}{T}}. \quad (4)$$

The received exponential law in equation (4) also has no singularity, and the derivations of DAL with parameters found from experimental $\tau_\alpha$ are far from the derivation of the experimental data in figures 2-4. Thereby, we conclude that both laws (VFT and DAL) are applicable only at $T < T_A$. In this sense it is not correct to compare VFT and DAL models by use the mean square errors for these functions including the temperature range $T > T_A$. The treatment by these functions in the range above $T_A$ will lead to incorrect optimization of the fitting parameters.

The importance of the choice of the temperature range can be demonstrated by estimation of the fitting quality of the different theoretical models. The comparison of accuracy of the fits by both VFT and DAL models was performed in [11] in terms of the quantity $\chi^2$. This parameter is defined by $\chi^2 = 1/(n-3)\sum[a_m - a_c]^2$ there $n$ is number of data points, $a_m$ and $a_c$ are the measured data and the calculated data, respectively [11]. We carried out the $\chi^2$ comparison of VFT and DAL models in case of propylene carbonate. The temperature dependence of the digitized dielectric relaxation data from [23] was fitted by the discussed models in two different temperature ranges. In the case of the full temperature range the ratio of $\chi^2_{VFT}/\chi^2_{DAL}$ is about 2.74. On the other hand, $\chi^2_{VFT}/\chi^2_{DAL}$ is equal 6.5 if the comparison is performed only at temperatures lower temperature $T_A$. Thus, the ratio of the $\chi^2_{VFT}/\chi^2_{DAL}$ depends significantly on the temperature range choice, and the exclusion of the range above $T_A$, where the models don't work, is important.

On other hand, from the general point of view it should exist a temperature above which the assumptions and approximations, working for dense and high-viscous matter, are not good. So, it is expected that above certain temperature the descriptions of liquid in terms of vacancies, free volume, network constraints and so on don't work. We believe that the transition at $T = T_A$ serves as a border above which the usual viscous liquid approximations fail.

In case of the FLD model the singularity, imbedded in equation (3) because of Heaviside function, could spread the applicability of this equation to the high-temperature part. Rather complicated functional form at $T < T^*$ changes to simple Arrhenius manner at $T > T^*$ where $\Theta(T^*-T)$ becomes to be zero. The derivative analysis of the FLD fit (found for the experimental data) is given by blue lines in figures 2-4. From these figures and table it is seen that the temperature $T^*$ differs remarkably from $T_A$ for all liquids considered.

One question arises. Maybe the difference between $T^*$ and $T_A$ is accidental, and the fit of the experimental data can be realized with $T^* = T_A$ without lowering the fit's quality? The results of the fits with free $T^*$ and $T^* = T_A$ in case of propylene carbonate is given in figure 5a. The quality of the fit becomes worse at fixed $T^* = T_A$. Figure 5b shows the difference between the experimental results and the fits by equation (3) with different $T^*$. Advantage of the case with the free value of $T^*$ is undoubted at this presentation. This means that the inconsistency between $T^*$ and $T_A$ needs a special analysis. We believe that this disagreement arises from the unsuccessful function for $\tau_\alpha(T)$ in equation (3) below $T_A$, which is compensated by the choice of $T^*$ in the numerical fit.

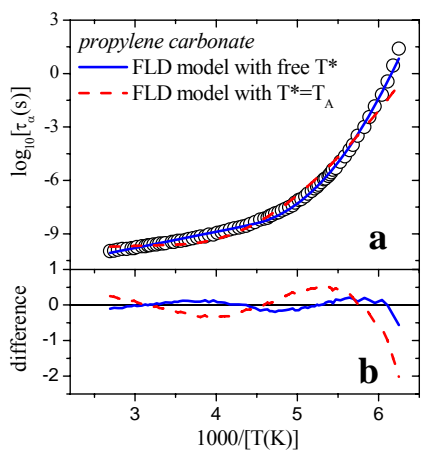

**Fig. 5** Results of fitting propylene carbonate's $\tau_\alpha$ by FLD model with free $T^*$ and $T^* = T_A$. Bottom part: difference between $\log\tau_\alpha$ and the fits.

To conclude, it was shown that both three-parametric model functions (VFT and DAL) don't describe a rather sharp transition from the Arrhenius to non-Arrhenius behaviour of $\tau_\alpha(T)$, which is clearly seen in the derivative analysis of three typical glass-formers. These functions should be applied and compared only at $T < T_A$, while the consideration also range $T > T_A$ can significantly distort conclusions about the preference of one or another models. The four-parametric FLD model indicates transition from Arrhenius to non-Arrhenius behaviour, but the transition temperature $T^*$ is systematically different from $T_A$ in cases considered. And we believe that this disagreement is due to inappropriate choice of $\tau_\alpha(T)$ dependence in the FLD model.

This work was supported by RFBR Grants No 09-03-00588, by Siberian Branch of RAS No 87, and by Ministry of Education and Science of RF, project No 2.1.1/1522.